\documentclass[12pt,letter]{article}
\pdfoutput=1
\usepackage{graphicx, epsfig, color,cite}
\usepackage{amsmath}
\usepackage{amssymb}
\usepackage{caption,subcaption,graphicx}
\usepackage[hidelinks]{hyperref}
\def\eslt{\not\!\!\!{E_T}}
\def\to{\rightarrow}

\def\bi{\begin{itemize}}
\def\ei{\end{itemize}}
\def\ta{\tilde a}
\def\te{\tilde e}

\def\tu{\tilde u}

\def\tb{\tilde b}

\def\tst{\tilde t}
\def\ttau{\tilde \tau}

\def\tg{\tilde g}
\def\tnu{\tilde\nu}
\def\tell{\tilde\ell}
\def\tq{\tilde q}

\def\tw{\widetilde\chi^{\pm}}
\def\tz{\widetilde\chi^0}
\def\alt{\lesssim}
\def\agt{\gtrsim}
\def\be{\begin{equation}}  
\def\ee{\end{equation}}  
\def\bea{\begin{eqnarray}}  
\def\eea{\end{eqnarray}}

\begin{document}
\begin{titlepage}
\begin{flushright}
OU-HEP-190501
\end{flushright}

\vspace{0.5cm}
\begin{center}
{\Large \bf Is the magnitude of the Peccei-Quinn scale\\ 
set by the landscape?
}\\ 
\vspace{1.2cm} \renewcommand{\thefootnote}{\fnsymbol{footnote}}
{\large Howard Baer$^1$\footnote[1]{Email: baer@ou.edu },
Vernon Barger$^2$\footnote[2]{Email: barger@pheno.wisc.edu},
Dibyashree Sengupta$^1$\footnote[3]{Email: Dibyashree.Sengupta-1@ou.edu}\\
Hasan Serce$^2$\footnote[4]{Email: serce@ou.edu},
Kuver Sinha$^1$\footnote[5]{Email: kuver.sinha@ou.edu} and
Robert Wiley Deal$^1$\footnote[6]{Email: rwileydeal@ou.edu}
}\\ 
\vspace{1.2cm} \renewcommand{\thefootnote}{\arabic{footnote}}
{\it 
$^1$Dept. of Physics and Astronomy,
University of Oklahoma, Norman, OK 73019, USA \\[3pt]
}
{\it 
$^2$Dept. of Physics,
University of Wisconsin, Madison, WI 53706 USA \\[3pt]
}

\end{center}

\vspace{0.5cm}
\begin{abstract}
\noindent
Rather general considerations  of the string theory landscape imply a
mild statistical draw towards large soft SUSY breaking terms tempered by the
requirement of proper electroweak symmetry breaking where SUSY contributions 
to the weak scale are not too far from $m_{weak}\sim 100$ GeV. 
Such a picture leads to the prediction that $m_h\simeq 125$ GeV while 
most sparticles are beyond current LHC reach. 
Here we explore the possibility that the magnitude of the 
Peccei-Quinn (PQ) scale $f_a$ is also set by string landscape
considerations within the framework of a compelling SUSY axion model.
First, we examine the case where the PQ symmetry arises as an 
accidental approximate global symmetry from a more fundamental
gravity-safe $\mathbb{Z}_{24}^R$ symmetry and where the 
SUSY $\mu$ parameter arises from a Kim-Nilles operator. 
The pull towards large soft terms then also pulls the PQ scale 
as large as possible.
Unless this is tempered by rather severe (unknown) cosmological or anthropic
bounds on the density of dark matter, then we would expect a far greater
abundance of dark matter than is observed. 
This conclusion cannot be negated by adopting a tiny axion misalignment 
angle $\theta_i$ because WIMPs are also overproduced at large $f_a$. 
Hence, we conclude that setting the PQ scale via anthropics is highly unlikely.
Instead, requiring soft SUSY breaking terms of order the gravity-mediation
scale $m_{3/2}\sim 10-100$ TeV  places the mixed axion-neutralino 
dark matter abundance into the intermediate scale sweet zone where 
$f_a\sim 10^{11}-10^{12}$ GeV. 
We compare our analysis to the more general case of a generic SUSY DFSZ 
axion model with uniform selection on $\theta_i$ but leading to the measured
dark matter abundance: this approach leads to a preference for 
$f_a\sim 10^{12}$ GeV.

\end{abstract}
\end{titlepage}

\section{Introduction}
\label{sec:intro}

The Standard Model (SM) is beset with several fine-tuning problems
that call for new physics beyond the SM. These include:
\begin{enumerate}
\item The gauge hierarchy problem\cite{Witten:1981nf} wherein the weak scale 
$m_{weak}\simeq m_{W,Z,h}\simeq 100$ GeV $\sim 10^{-14}m_{\rm GUT}$ (where
$m_{\rm GUT}\simeq 2\times 10^{16}$ GeV).
\item The cosmological constant (CC) problem\cite{CC} 
wherein the measured value of the 
cosmological constant $\Lambda\simeq 10^{-120}m_P^4$ 
(where naively it is expected that $\Lambda\simeq m_P^4$ with
$m_P$ being the reduced Planck mass).
\item The strong CP problem\cite{SCP} wherein the $\bar{\theta}$ 
coefficient of the CP-violating
$\bar{\theta}(g_s^2/32\pi )G_{\mu\nu A}\tilde{G}_A^{\mu\nu}$ QCD Lagrangian term is $\alt 10^{-10}$ 
times its expected value
from the 't Hooft theta vacuum solution to the $U(1)_A$ problem\cite{U1A}.
\end{enumerate}

The most elegant solution to problem \#1 is to extend the Poincare' 
group of spacetime symmetries to include weak scale broken supersymmetry 
(WSS)\cite{DG,wss} 
wherein all quadratic divergences to $m_h^2$ necessarily cancel. 
WSS is in fact supported by {\it four} sets of measurements: 
\begin{itemize}
\item The measured values of the gauge couplings actually unify under 
Minimal Supersymmetric Standard Model (MSSM) evolution 
whereas they do not under SM evolution\cite{gauge}.
\item The measured value of the top quark mass is just right to generate 
a radiative breakdown of EW symmetry in the MSSM (assumed valid up to scales
$Q\simeq m_{\rm GUT}$)\cite{rewsb}.
\item The measured value of $m_h\simeq 125$ GeV falls squarely within the 
narrow window of MSSM required values where $m_h\alt 135$ GeV\cite{mhiggs} and 
\item  the measured value of $m_W\simeq 80.4$ GeV favors heavy WSS over the SM
assuming the measured value of $m_t\simeq 173.2$ GeV\cite{sven2006}.
\end{itemize}
Early concern over the non-appearance of SUSY particles at LHC Run 2 has been 
allayed by renewed scrutiny of naturalness measures\cite{ltr,rns}: 
updated analyses now require only $m_{\tg}\alt 6$ TeV and $m_{\tst_1}\alt 3$ TeV\cite{rns,upper,jamie} as compared to present LHC limits that 
$m_{\tg}\agt 2$ TeV and $m_{\tst_1}\agt 1$ TeV. 
In the MSSM, only higgsinos are required by naturalness to have 
weak scale masses, and these particles
are very difficult (but not impossible) to see at LHC\cite{SDLJMET,rns@lhc}. 

Some understanding of the magnitude of the cosmological constant has emerged
from the string theory landscape of vacua solutions\cite{landscape}. 
In a discretuum\cite{Bousso:2000xa} of vacua states with all possible values 
for $\Lambda: -m_P^4\to +m_P^4$, then one 
expects $\Lambda$ as large as possible subject to the requirement that
galaxies be able to condense, which is a seeming precondition for a
pocket universe containing sentient observers. 
Indeed, Weinberg\cite{Weinberg:1987dv} used such reasoning to predict 
the value of $\Lambda$  to within a factor of several of its value 
which was measured more than a decade later. 
Such an anthropic solution to the CC problem 
emerges naturally from string theory incuding a vast landscape of flux 
vacua, estimated as $\sim 10^{500}$ such possibilities\cite{doug}.

The most elegant solution to the strong CP problem involves the introduction 
of a new global $U(1)$ PQ symmetry\cite{pq} which is spontaneously broken (SSB) at
some scale $f_a\sim 10^9-10^{16}$ GeV.\footnote{In accord with the PDG\cite{pdg},
we take $f_A\equiv f_a/N_{\rm DW}$ where $N_{\rm DW}$ is the domain-wall number which 
is $N_{\rm DW}=6$ for the DFSZ axion model assumed here.} 
The (pseudo-)Goldstone boson which 
emerges from SSB, the axion $a$\cite{ww}, allows for a dynamical relaxation of
the $G\tilde{G}$ QCD Lagrangian terms to zero thus solving the strong CP
problem. A remnant of the PQ procedure is the existence of a physical
axion particle which also turns out to be a solid candidate for
cold dark matter (CDM) in the universe\cite{aCDM}. 
Remnant axion CDM is being searched for at a variety of experiments, 
the most sensitive of which are the microwave cavity searches\cite{admx}. 
While the PQ axion solution to the strong CP problem is indeed compelling, 
it is beset by two problems of its own.
\begin{itemize}
\item Global symmetries are not respected by gravitational interactions and thus
the PQ symmetry is not expected to be fundamental\cite{KMR,grav}. 
Instead, PQ symmetry is expected to emerge as an accidental, 
approximate symmetry from some more fundamental gravity-safe symmetry. 
To be gravity-safe, the resulting PQ symmetry must be of exceptionally 
high quality: the PQ breaking contributions to the 
axionic potential must be suppressed by at least eight powers of 
$m_P$\cite{KMR} in order for $\bar{\theta}\alt 10^{-10}$: 
$V_{PQB}\sim \phi^{10}/m_P^8$ (where $\phi$ stands for generic scalar fields). 
This is a very high bar to hurdle!
\item In string theory, many candidate axions can emerge, but with a PQ scale
$f_a\sim m_{\rm GUT}$ to $m_{\rm string}$\cite{Choi:1985je,Svrcek:2006yi}. 
Meanwhile, cosmological (dark matter) constraints
seem to require $f_a\sim 10^{11}-10^{12}$ GeV\cite{aCDM}. 
A further problem then is:
what accounts for the apparent suppression of the PQ breaking scale?
\end{itemize}

The goal of this paper is to examine the second of these axionic concerns 
in the context of the string theory landscape: can the magnitude of the 
PQ breaking scale be understood from landscape considerations within a 
well-motivated model for axion (and WIMP) dark matter?
A number of previous works have also addressed this question and these will
be briefly reviewed in Subsec.~\ref{ssec:review}.
The first of the PQ concerns--gravity safety-- was recently addressed 
within the context of anomaly-free (up to a Green-Schwarz term) 
discrete $R$ symmetries which forbid the SUSY $\mu$ term while 
respecting grand unification conditions\cite{lrrrssv}.
It was found in Ref.~\cite{bbs} that two closely related SUSY axion models-- 
dubbed hybrid CCK\cite{cck} (hyCCK) and hybrid SPM\cite{spm} (hySPM)-- 
were found to be gravity-safe under a $\mathbb{Z}_{24}^R$ discrete $R$-symmetry 
which also led to 
\begin{itemize}
\item suppression of the SUSY $\mu$ term, 
\item suppression of the various renormalizable $R$-parity violating terms and 
\item suppression of the dangerous dimension-5 proton decay operators, 
\end{itemize}
all while allowing for see-saw neutrinos. 
In such a setting, the $\mathbb{Z}_{24}^R$ symmetry
(and resulting accidental approximate $U(1)_{PQ}$ symmetry) are broken as a
consequence of SUSY breaking. Thus, the PQ scale emerges as a derived value 
depending on the soft SUSY breaking terms. We will assume the hyCCK model
in Sec. \ref{sec:Z24R} and derive a probability distribution for the 
resulting magnitude of the PQ scale $f_a$ for an assumed upper bound on the
allowed dark matter abundance in pocket universes.
Requiring a (pocket) universe without overproduction of mixed axion-WIMP
dark matter by a (arbitrary) factor four excess beyond its measured value 
then leads to a most probable value of $f_a\sim 10^{14}$ GeV.
We also derive a probability distribution for the initial axion 
misalignment angle $\theta_i$ which tends to favor smaller values of 
its expected range. Typically this leads to a large overproduction
of dark matter beyond its observed value.
If one assumes axion-only dark matter, then one may compensate for the
large value of $f_a$ by allowing for a small value of the initial 
axion misalignment angle $\theta_i\sim 0$\cite{Linde:1987bx}. In our approach, where
SUSY stabilizes the weak scale and the $\mathbb{Z}_{24}^R$ symmetry 
yields gravity-safety, then WIMPs are also overproduced and a tiny value
of $\theta_i$ cannot save the day for $f_a \agt 10^{14}$ GeV.


Some motivation for our approach comes from earlier analyses by
Douglas which explored a statistical approach to the magnitude of the
SUSY breaking scale. In Ref's \cite{Douglas:2004qg} and \cite{Douglas:2004zg}, 
it is assumed that all real-valued SUSY breaking scales are equally 
likely in a fertile patch of the landscape which contains the MSSM as the low
energy effective theory.
In the case of $F$-term SUSY breaking, since the $F$ terms are complex-valued
fields and the magnitude of SUSY breaking depends on the modulus of 
$\langle F\rangle$, then one expects the magnitude of soft terms
$m_{soft}\sim m_{3/2}\sim m_{hidden}^2/m_P$ to enjoy a {\it linearly increasing}
statistical distribution in the landscape. 
For a variety of hidden sector $F$ and $D$-term fields contributing to 
SUSY breaking, then one expects instead $m_{soft}^n$ where 
$n=2n_F+n_D-1$ and where $n_F$ is the number of contributing 
$F$-term SUSY breaking fields and $n_D$ is the number of $D$-term SUSY 
breaking fields.

Naively, the increasing soft term prior would suggest soft terms most probably 
at energy scales far beyond the weak scale. 
However, Agrawal {\it et al.}\cite{Agrawal:1998xa} have computed that 
if the weak scale is increasing by a factor of 2-5 beyond its measured value,
 then nuclear physics is modified in ways that are unlikely to lead to a 
livable universe (as we understand it).
Requiring SUSY contributions to the weak scale to be within a factor of a few
(four) of its measured value is the same as requiring the naturalness 
measure\cite{ltr,rns} $\Delta_{\rm EW}\alt 30$\cite{Baer:2016lpj}. 
By requiring that the weak scalar potential is properly
broken to $SU(3)_C\times U(1)_{EM}$ and no contribution to the weak scale
exceeds a factor four beyond the weak scale (in the presence of a 
natural value of $\mu\sim 200$ GeV which emerges from our assumed
solution to the SUSY $\mu$ problem),
then the distribution of soft terms becomes bounded from above.
For the case of a mild $n=1$ statistical draw on soft terms, then there is 
large mixing in the top-squark mass matrix leading to enhanced Higgs mass 
radiative corrections and hence $m_h\simeq 125$ GeV whilst
the gluinos and squarks are pulled beyond LHC search 
limits\cite{Baer:2017uvn,Baer:2019xww}. 
An exception is the SUSY-preserving $\mu$ term which directly contributes to 
the magnitude of the weak scale and thus must be of order $\mu\sim 100-350$ GeV.
The $\mu$ parameter leads to a rather light set of four higgsinos whose 
parameter space is only now beginning to be explored at LHC 
via the soft dilepton plus jet plus $\eslt$ signature\cite{SDLJMET}. 
Thus, in the SUSY landscape picture with a mild
statistical draw towards large soft terms, the prediction is that the LHC will
see exactly that which it does see: a light Higgs of mass $m_h\simeq 125$ GeV
with as yet no sign of sparticles\cite{Baer:2019xww}.

\subsection{Review of some previous work}
\label{ssec:review}

Here, we briefly review some previous related works to give context
for our directions.
\begin{itemize}
\item In Ref.~\cite{Linde:1987bx}, Linde assumes a scenario where 
inflation continues past the PQ phase transition (which alleviates the
axion domain wall problem) giving rise to a uniform axion field strength
$a\equiv \theta_i f_a$ throughout the observable universe. 
He argues that an increased value of initial axion field strength 
by a factor ten in pocket universes leads 
to an increase in matter content of galaxies by a factor $\sim 10^8$ 
likely leading to conditions hostile to life as we know it. 
Since $\Omega_a h^2\sim \theta_i^2 f_a^{7/6}$, then small values of 
$\theta_i$ are favored and hence values of $f_a\gg 10^{12}$ GeV 
can be accommodated.
\item In Hellerman and Walcher Ref.~\cite{Hellerman:2005yi}, 
the authors examine axion CDM 
and galaxy formation in a multiverse setting where all other parameters of
$\Lambda$CDM models are fixed at the values of our pocket universe. 
By requiring structure formation 
before the onset of cosmological constant domination, the authors derive upper
bounds on the ratio of dark matter-to-baryons $\rho_{DM}/\rho_B$ which are
typically of order $10^4-10^5$ while our universe sits at the lower edge of 
allowed values $\sim 5$. They conclude that anthropic constraints are 
{\it unlikely} to explain our observed value of $\rho_{DM}/\rho_B\sim 5$.
\item In Wilczek and in Tegmark {\it et al.} 
Ref's.~\cite{Wilczek:2004cr,Tegmark:2005dy}, a model with axion-only CDM
is assumed where PQ breaking occurs before the end of inflation. For
uniform sampling of $\theta_i$, then the prior distribution of axion CDM is
the mild distribution $f(\rho_{\rm CDM})\sim 1/\sqrt{\rho_{\rm CDM}}$. 
Imposing anthropic limits from galaxy formation, star and black hole 
formation, solar system stability and stiff upper limits on the 
density of habitable halos, then it is found that the measured 
DM density is comparable to expectations from the multiverse.
\item Freivogel in Ref.~\cite{Freivogel:2008qc} adopts the same axion-only 
CDM model as in Ref~\cite{Tegmark:2005dy} but then uses Bousso's causal 
diamond measure to provide selection constraints. 
He fixes $\rho_B/\rho_{\gamma}$ to its observed value, fixes $f_a\gg 10^{12}$ GeV 
and allows only the misalignment angle $\theta_i$ to vary uniformly and 
calculates the probability of observing a particular value of  
$\xi =\rho_{\rm DM}/\rho_B$ which is allowed to vary. 
He finds 68\% of observers see $\xi\le 15$ and 95\% of observers see 
$\xi\le 65$. In our pocket universe with $\xi\sim 5$, he concludes that 
the observed value of $\xi$ is then reasonable. 
However, it is not completely improbable to observe $\xi$ values different 
from our universe.
\end{itemize}
Some related papers\cite{Bousso:2009ks,Bousso:2010zi,Bousso:2013rda,DEramo:2014urw} argue that comparable values 
for $\Omega_\Lambda$ and $\Omega_{\rm DM}$ should emerge from the multiverse.

In these works, a particularly parsimonious approach-- 
of the SM amended with a simple axionic extension-- is adopted. 
While admittedly simple, it contains two problems which make them 
likely unrealistic. 
First, there is nothing in them to stabilize the weak scale, and thus 
we would expect a value for the weak scale far beyond its measured value.
The inclusion of WSS solves this problem. Second, it is well known that
the global $U(1)_{PQ}$ needed for an axionic solution to the strong CP
problem is intrinsically incompatible with the inclusion of gravitation
and with embedding in string theory in particular. In the following, we 
adopt a particular SUSY axion model based upon a more fundamental
discrete $R$-symmetry which may arise from compactification of extra
space dimensions in string theory. For a strong enough discrete symmetry,
in this case $\mathbb{Z}_{24}^R$, then the emerging accidental, approximate
global PQ symmetry is sharp enough such as to allow for the PQ solution 
to the strong CP problem.

\section{Probability distributions for the PQ scale and 
$\theta_i$ from a SUSY axion model based upon 
a gravity-safe $\mathbb{Z}_{24}^R$ symmetry}
\label{sec:Z24R}

In this Section, we first introduce the gravity-safe hyCCK SUSY axion model
where the global PQ symmetry emerges as an accidental, approximate 
global symmetry from a more fundamental $\mathbb{Z}_{24}^R$ symmetry.
Next, we review calculation of the mixed axion-WIMP dark matter abundance
using eight coupled Boltzmann equations. 
Then, we combine statistical selection of the PQ scale with the 
requirement against overproduction of dark matter within pocket universes 
which populate an eternally-inflating\cite{Guth:2007ng}
multiverse to derive probability distributions for the magnitude of the
PQ energy scale and also for the initial axion misalignment angle $\theta_i$.

\subsection{MSSM augmented by gravity-safe PQ sector based on $\mathbb{Z}_{24}^R$ discrete $R$-symmetry}
\label{ssec:hyCCK}

In Ref. \cite{Baer:2017uvn}, probability distributions for Higgs and sparticle
masses were derived from the string landscape assuming 
Douglas'\cite{Douglas:2004qg,Douglas:2004zg} 
$m_{soft}^n$ (with $n=0,1,2$) soft term prior 
coupled with a veto $\Theta (30-\Delta_{\rm EW})$ on non-standard
vacua or too large of SUSY contributions to the weak scale. 
The $n=1$ sampling generates a probability distribution for $m_h$ which 
peaked at $m_h\simeq 125$ GeV while sparticle mass predictions were 
characterized by
\begin{enumerate}
\item $m_{\tg}\sim 4\pm 2$ TeV,
\item $m_{\tst_1}\sim 1.5\pm 0.5$ TeV,
\item $m_A\sim 3\pm 2$ TeV,
\item $m_{\tw_1,\tz_{1,2}}\sim 200\pm 100$ GeV and
\item $m_{\tq,\tell}\sim 20\pm 10$ TeV.
\end{enumerate}
In accord with these results, we will adopt for illustrative purposes a
SUSY benchmark point from the NUHM3 model\cite{nuhm2} with parameters
$m_0(1,2)=16$ TeV, $m_0(3)=5$ TeV, $m_{1/2}=1.5$ TeV, $A_0=-7$ TeV, 
$\tan\beta =10$ with $\mu =200$ GeV and $m_A=3$ TeV. 
We generate sparticle mass spectra using Isajet 7.88\cite{isajet}
and find the spectra provided in Table \ref{tab:bm} and labelled as
landSUSY (landscape SUSY). 
The spectra lie well within the landscape SUSY predictions for an 
$n=1$ mild draw to large soft terms\cite{Baer:2017uvn}.
Our final results will hardly depend on significant deviations from the 
landSUSY benchmark model which mainly sets a natural value for the 
$\mu$ parameter and the saxion and axino branching fractions which
enter into the relic density calculation.
%
\begin{table}\centering
\begin{tabular}{lc}
\hline
parameter & landSUSY \\
\hline
$m_0(1,2)$      & $16$ TeV \\
$m_0(3)$       & $5$ TeV \\
$m_{1/2}$      & 1.5 TeV \\
$A_0$      & -7 TeV \\
$\tan\beta$    & 10  \\
\hline
$\mu$          & 0.2 TeV  \\
$m_A$          & 3 TeV \\
\hline
$m_{\tg}$   & 3619 GeV \\
$m_{\tu_L}$ & 16211 GeV \\
$m_{\tu_R}$ & 16264 GeV \\
$m_{\te_R}$ & 15956 GeV \\
$m_{\tst_1}$& 1294 GeV \\
$m_{\tst_2}$& 3561 GeV \\
$m_{\tb_1}$ & 3605 GeV \\
$m_{\tb_2}$ & 4999 GeV \\
$m_{\ttau_1}$ & 4749 GeV \\
$m_{\ttau_2}$ & 4982 GeV \\
$m_{\tnu_{\tau}}$ & 4951 GeV \\
$m_{\tw_1}$ & 210 GeV \\
$m_{\tw_2}$ & 1312 GeV \\
$m_{\tz_1}$ & 200 GeV \\ 
$m_{\tz_2}$ & 207 GeV \\ 
$m_{\tz_3}$ & 688 GeV \\ 
$m_{\tz_4}$ & 1320 GeV \\ 
$m_h$       & 125 GeV \\ 
\hline
$\Omega_{\tz_1}^{std}h^2$ & 0.01 \\
$BF(b\to s\gamma)\times 10^4$ & $3.0$ \\
$BF(B_s\to \mu^+\mu^-)\times 10^9$ & $3.8$ \\
$\sigma^{SI}(\tz_1, p)$ (pb) & $1.0\times 10^{-9}$ \\
$\sigma^{SD}(\tz_1 p)$ (pb)  & $2.0\times 10^{-5}$ \\
$\langle\sigma v\rangle |_{v\to 0}$  (cm$^3$/sec)  & $2\times 10^{-25}$ \\
$\Delta_{\rm EW}$ & 23.3 \\
\hline
\end{tabular}
\caption{Input parameters (TeV) and masses (GeV)
for a landscape SUSY benchmark point from the NUHM3 model
with $m_t=173.2$ GeV using Isajet 7.88\cite{isajet}.
}
\label{tab:bm}
\end{table}

We will augment the landscape SUSY spectra with a PQ sector from the
hybrid CCK\cite{cck} model where the superpotential is given by
\begin{eqnarray}
W_{hyCCK}&\ni &f_u QH_uU^c +f_dQH_dD^c+f_\ell LH_dE^c+f_\nu LH_uN^c\nonumber \\
&+& fX^3Y/m_P+\lambda_\mu X^2H_uH_d/m_P+M_NN^cN^c/2
\label{eq:WhyCCK}
\end{eqnarray}
where we have introduced MSSM singlet fields $X$ and $Y$
with PQ charges listed in Table \ref{tab:hyCCK}. 
The hyCCK model is thus a particular example of a SUSY DFSZ axion model where 
PQ field $X$ couples to the two Higgs doublets thus providing a 
solution to the SUSY $\mu$ problem\footnote{When the PQ field $Y$ couples 
to the two Higgs doublets, the resulting model is hySPM model which 
is also an example of a gravity-safe model that solves strong CP problem 
and the SUSY $\mu$ problem. This model gives $f_a$ values 
similar to those obtained in the hyCCK model\cite{bbs}.}.
The model assumes an 
underlying $\mathbb{Z}_{24}^R$ discrete $R$ symmetry which forbids 
1. renormalizable RPV terms, 2. the usual $\mu$ term and 
3. dimension-5 proton decay operators while allowing 
the required Yukawa couplings and see-saw neutrino terms\cite{lrrrssv}.
\begin{table}[!htb]
\renewcommand{\arraystretch}{1.2}
\begin{center}
\begin{tabular}{c|cccccccccc}
multiplet & $H_u$ & $H_d$ & $Q_i$ & $L_i$ & $U_i^c$ & $D_i^c$ & $E_i^c$ & $N_i^c$ & X & Y \\
\hline
 $\mathbb{Z}_{24}^R$ charge & 16 & 12  & 5 & 9 & 5 & 9 & 5 & 1 & -1 & 5 \\
\hline
PQ charge & -1 & -1  & 1 & 1 & 0 & 0 & 0 & 0 & 1 & -3 \\
\hline
\end{tabular}
\caption{ $\mathbb{Z}_{24}^R$ and PQ charge assignments for various superfields of the 
hyCCK model.
}
\label{tab:hyCCK}
\end{center}
\end{table} 
The lowest order PQ breaking superpotential terms are 
$X^8Y^2/m_P^7$, $Y^{10}/m_P^7$ and $X^4Y^6/m_P^7$ leading to lowest order
scalar potential terms suppressed by powers of $m_P^8$: thus, the 
underlying $\mathbb{Z}_{24}^R$ symmetry renders the model gravity-safe
according to the KM-R requirements\cite{KMR}.
The required $U(1)_{PQ}$ symmetry arises as an accidental approximate
global symmetry which arises from an underlying more fundamental
discrete $\mathbb{Z}_{24}^R$ symmetry.

The scalar potential for hyCCK, augmented with the corresponding soft SUSY breaking terms, is given by
\begin{equation}
V_{hyCCK}\ni \frac{f^2}{m_P^2}\left[9|\phi_X|^4|\phi_Y|^2+|\phi_X|^6\right]+m_X^2|\phi_X|^2+m_Y^2|\phi_Y|^2+(f A_f\phi_X^3\phi_Y/m_P+h.c.) .
\label{eq:VhyCCK}
\end{equation}

Minimization conditions for the hyCCK model can be found in Ref. \cite{radpq}.
The scalar potential develops a non-zero minimum at 
$\langle\phi_X\rangle\equiv v_X$ and $\langle\phi_Y\rangle\equiv v_Y$ for a
sufficiently large soft term $-A_f$, thus breaking the underlying 
$\mathbb{Z}_{24}^R$ and accidental, approximate $PQ$ symmetries. 
The PQ breaking vev is given by $v_{PQ}=\sqrt{v_X^2+9v_Y^2}$ where then
$f_a=\sqrt{2}v_{PQ}$. 
In accord with expectations from supergravity models, 
we will assume $m_X=m_Y=m_{\ta}=m_s\equiv m_{3/2}$\cite{mass}.
Thus, in this model, the PQ scale is a derived consequence of SUSY breaking. 

\begin{figure}[tbp]
\begin{center}
\includegraphics[height=0.4\textheight]{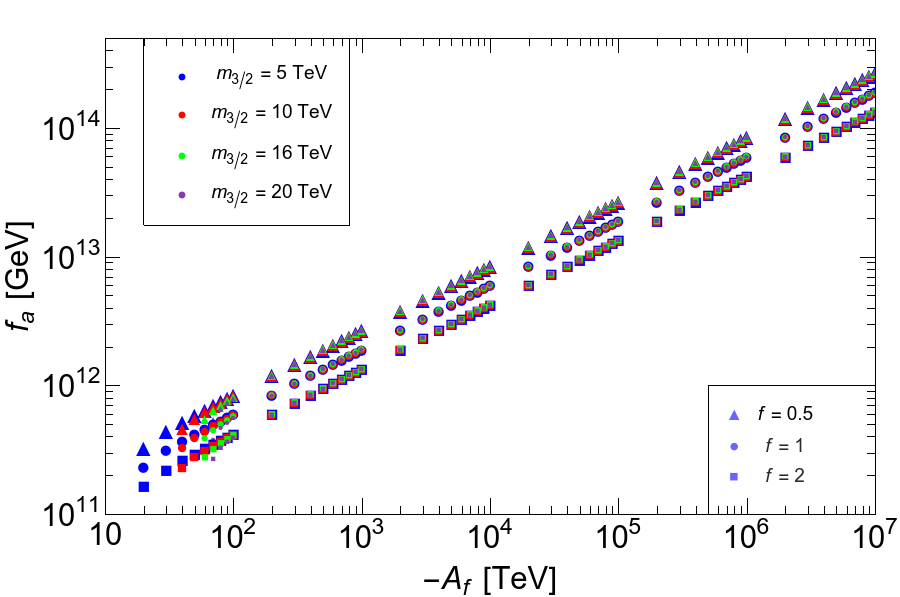}
\caption{
Value of Peccei-Quinn scale $f_a$ vs. hyCCK soft parameter $-A_f$
for various values of $m_X=m_Y\equiv m_{3/2}$ and three different 
values of $f$.
\label{fig:fa}}
\end{center}
\end{figure}

The calculated value of $f_a$ is given in Fig. \ref{fig:fa} 
as a function of the $-A_f$ soft term assuming various values of 
$m_X=m_Y=m_0(1,2)\equiv m_{3/2}$ and three different 
values of $f$.
From Fig.~\ref{fig:fa}, we see that $f_a$ has a monotonically increasing value
with increasing $|-A_f|$. For a particular value of  $-A_f$ and 
$m_X=m_Y=m_0(1,2)\equiv m_{3/2}$ if the value of $f$ is reduced 
by a factor of 2, then $f_a$ increases by approximately $41\%$ and
if the value of $f$ is increased 
by a factor of 2, then $f_a$ decreases by approximately $41\%$.   
Since $-A_f$ doesn't contribute directly to the 
determination of the weak scale, then there is no (anthropic) 
upper bound on its value and one might expect $-A_f$ and hence 
$f_a$ to lie far beyond the well-known cosmological sweet spot where 
$f_a\sim 10^{11}-10^{12}$ GeV\cite{aCDM}.   

\subsection{Relic density of mixed axion-WIMP dark matter}
\label{ssec:az1}

The evaluation of mixed axion-WIMP dark matter from SUSY axion models is
more complicated than simply adding the WIMP thermal abundance to
the coherent-oscillation-produced axions.
To evaluate the mixed neutralino-axion relic density, we apply the
eight-coupled-Boltzmann equation computer code developed in 
Ref's~\cite{dfsz2,andre}. 
For brevity, we will not reproduce here the eight coupled Boltzmann 
equations which can instead be found in Ref. \cite{dfsz2}.
The code relies on the IsaReD\cite{isared} 
calculation of $\langle\sigma v\rangle (T)$ which is a crucial input 
to the coupled Boltzmann calculation.
Starting from the time of re-heat with temperature $T_R$
at the end of the inflationary epoch, 
the computer code tracks the coupled abundances
of radiation ({\it i.e.} SM particles), neutralinos, axinos, gravitinos, 
saxions and axions (the latter two consists of both thermal/decay-produced 
and coherent oscillation-produced (CO) components).

The CO-produced abundance of axions is determined in part by the
axion field initial misalignment angle $\theta_i$~\cite{aCDM,vg1}.
For numerical analyses, we adopt a simple formula
\be
\Omega_a^{\rm CO}h^2\simeq 0.23 f(\theta_i )\theta_i^2
\left(\frac{f_a/N_{\rm DW}}{10^{12}\ {\rm GeV}}\right)^{7/6}
\label{eq:axionCO}
\ee 
where $f(\theta_i )=\left[\log \left( e/(1-\theta_i^2/\pi^2)\right)\right]^{7/6}$
is the anharmonicity factor~\cite{vg1} and $N_{\rm DW}$ is the domain wall number
($=6$ for the DFSZ model).
In previous work the initial misalignment angle $\theta_i$ is adjusted to
gain the measured value of the relic abundance. In the current work, we
will allow for a uniform distribution of $\theta_i: 0-\pi$ values 
since we are scanning over many pocket universes
which arise as subuniverses of the more vast multiverse.
\begin{figure}[tbp]
\begin{center}
\includegraphics[height=0.35\textheight]{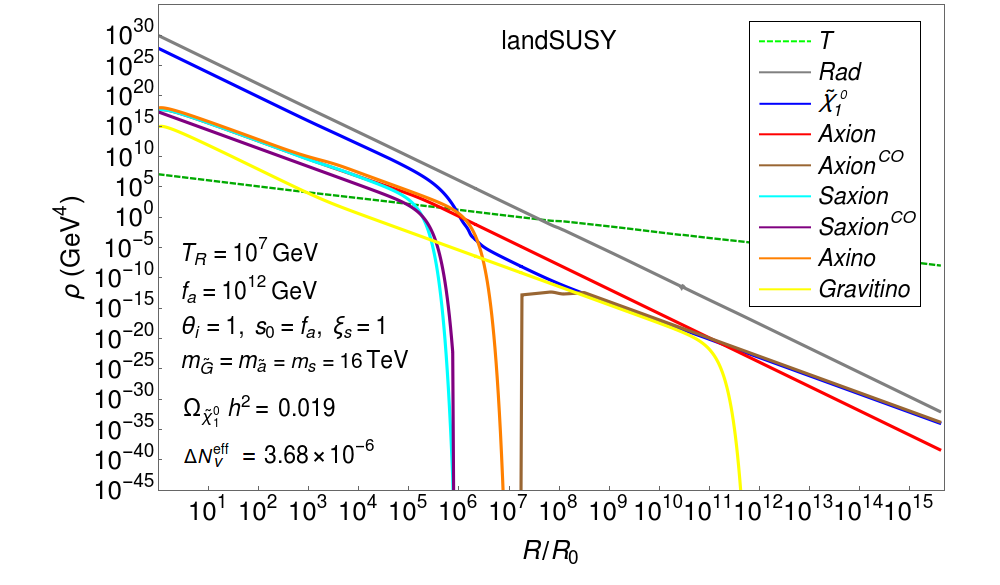}\\
\caption{A plot of various energy densities $\rho$ vs. 
scale factor $R/R_0$ starting from $T_R=10^7$ GeV until the 
era of entropy conservation from our eight-coupled Boltzmann 
equation solution to the mixed axion-neutralino relic density 
in the SUSY DFSZ model for the landscape SUSY benchmark point. 
We take $\xi_s=1$.
The corresponding temperature $T$ is denoted by the dashed green line
where in this case the $y$-axis is interpreted as $T$ in GeV.
\label{fig:rhoT}}
\end{center}
\end{figure}

In Fig. \ref{fig:rhoT} we show the energy densities of various species
vs. scale factor $R/R_0$ that influence the ultimate dark matter abundance 
for the landscape SUSY benchmark point landSUSY in Table \ref{tab:bm}. 
Here, $R_0$ is the reference scale factor at the beginning of re-heat 
and the corresponding temperature $T$ is shown by the dashed green line
(where instead the $y$-axis is interpreted as temperature in GeV).
We take $T_R=10^7$ GeV \footnote{This value of $T_R$
is in accord with well-motivated baryogenesis mechanisms such as 
non-thermal or Affleck-Dine leptogenesis~\cite{lepto}.} 
and $f_a=s_0=10^{12}$ GeV and where $s_0$ denotes the initial saxion field 
value. We also take $m_{\ta}=m_s=m_{3/2}=16$ TeV.
The blue curve denotes the neutralino abundance which freezes out at
$R\sim 10^6 R_0$ or $T\sim 10$ GeV. 
The saxion and axion contributions are split into their thermally- and 
decay-produced components and their coherent-oscillation (CO) produced
components.
Saxions decay around $R\sim 10^5R_0$ ($T\sim 10$ GeV) whilst axinos 
decay around $R\sim 10^6R_0$ ( or $T\sim 1$ GeV). 
The saxion decays depend on a model dependent coupling $\xi_s$ 
which governs the saxion decay rate $s\to aa$ and $s\to\ta\ta$\cite{dfsz2}.
We take $\xi_s=1$ so these decays are turned on. 
(Of course, for our case the $s\to\ta\ta$ decay is not kinematically open
so $s$ decays mainly to $aa$ but also to other MSSM particles).
CO-produced axions (brown curve) start to oscillate around $T\sim 1$ GeV
and become the dominant component of dark matter as one enters the 
era of entropy conservation on the right-hand-side of the plot. 
Due to late decays of axinos, which occur after neutralino freeze-out, 
the neutralino abundance increases to $\Omega_{\tz_1}h^2 \simeq 0.02.$

\begin{figure}[tbp]
\begin{center}
\includegraphics[height=0.35\textheight]{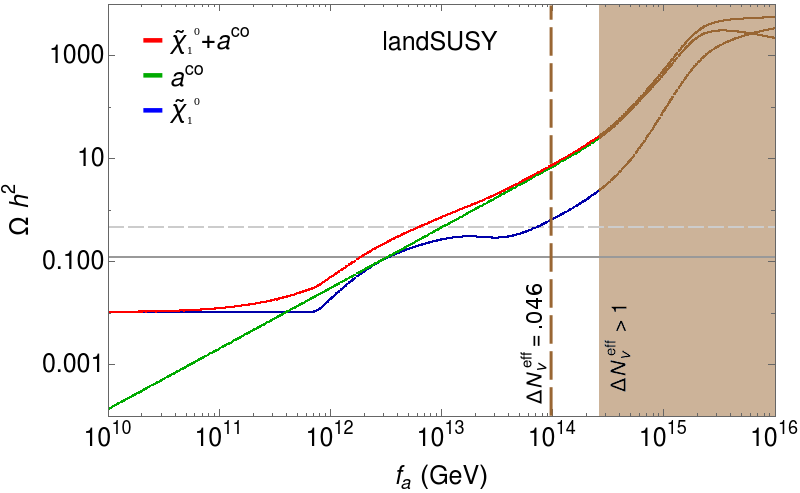}
\caption{
Relic density of axion and higgsino-like WIMP DM versus $f_a$
for the landSUSY benchmark point with $\theta_i=1$.
The red curve denotes the sum of axion plus WIMP dark matter
while green denotes the separate axion abundance and the blue curve denotes the separate WIMP abundance. 
The curves become brown when $\Delta N_\nu^{eff}>1$.
\label{fig:Oh2}}
\end{center}
\end{figure}

To gain some perspective on the expected relative abundances of
mixed axion-WIMP dark  matter, 
in Fig. \ref{fig:Oh2} we show the relic density of mixed axion-WIMP
dark matter vs. $f_a$ for the landSUSY benchmark point and with $T_R=10^7$ GeV
and $m_s=m_{\ta}=m_{3/2}=16$ TeV and where 
$\theta_i=\theta_s=1$\footnote{Here, the saxion field strength 
$s =\theta_s.f_{a}$.}. 
The green curve corresponds to the axion relic density while the blue curve 
corresponds to the WIMP relic density. 
The red curve shows the total relic density. 
We see that for low values of $f_a$, the axion relic density-- 
arising here from coherent oscillations corresponding to Eq. \ref{eq:axionCO}--
is highly suppressed. 
Also, the thermally-produced WIMP dark matter
is highly suppressed due to the higgsino-like nature of the LSP
which enhances its annihilation rate. 
The WIMP relic density is also highly suppressed by co-annihilations 
with the slightly heavier higgsinos
$\tw_1$ and $\tz_2$. Thus, for $f_a\sim 10^{10}-10^{12}$ GeV, we expect
typically an under-production of mixed axion-higgsino DM.
As $f_a$ increases, the CO-produced axions steadily increase while
WIMPs remain at their thermally-produced level. By $f_a\sim 10^{12}$ GeV, 
the axino and saxion decay rates are sufficiently suppressed 
(by $\Gamma_{\ta,s}\sim 1/f_a^2$) that they begin
decaying into higgsinos {\it after} WIMP freeze-out, thus augmenting 
the WIMP abundance with a non-thermal, decay-produced component. 
By $f_a\sim 3\times 10^{12}$ GeV, then the mixed axion-WIMP abundance saturates
the measured value $\Omega_{\rm CDM}h^2\simeq 0.12$, and where at this point 
CDM consists nearly equally of  axions along with a comparable
thermal and non-thermal WIMP component. 
In this region, the non-thermal WIMP component arises mainly from thermal
axino production followed by late $\ta$ decays in the early universe.
As $f_a$ increases further, the thermal axino production rate falls off
rapidly so that the WIMP abundance levels off. For even higher values of 
$f_a\agt10^{14}$ GeV, saxion production via COs becomes large and so 
saxion-decay produced WIMP production rapidly rises. 
In addition, the $s\to aa$ decays sharply increase the already 
over-produced axions. 
These relativistic axions also lead to violation of limits on 
relativistic species present in the early universe characterized in terms of 
the effective number of neutrinos parameter $\Delta N_{\nu}^{eff}$ which we take
(very conservatively) to be $\alt 1$ (brown curve).\footnote{
In the Particle Data Book\cite{pdg}, it is tabulated that $N_{eff}=3.13\pm 0.32$.}
For $f_a\agt2\times 10^{15}$ GeV, then entropy dilution of all relics 
from CO-produced saxions can suppress the mixed axion-neutralino
relic abundance.

In terms of the string theory landscape, we see that allowing 
values of $f_a\sim m_{\rm GUT}$ could lead to dark matter overproduction
by a factor of $\sim 10^4$ compared to its measured value.
As noted by Linde and others, it might be hard to visualize 
the existence of observers in a universe with such an overabundance
of dark matter.
Precisely how much of an overabundance of dark matter is 
anthropically too much is an open question. 
But clearly, if such a limit exists, then it would place an upper limit
on the value of $f_a$. Even requiring a modest factor of four overabundance, 
indicated by the dashed gray horizontal line, would already require
a value $f_a\alt 10^{13}$ GeV.
This upper bound is well below the expected
magnitude for $f_a$ from string theory where instead 
$f_a\sim 10^{16}-10^{18}$ GeV is typically expected\cite{sw}. 
The bound on $f_a$ from the axion abundance may be considered a softer bound
since it is possible to lower the axion abundance with a smaller value of
$\theta_i\sim 0$ (although if $\theta_i$ scans on the landscape, then
$\theta_i\sim 1$ is to be expected). However, we see that a bound on $f_a$
still obtains from the WIMP contribution to $\Omega_{a\tz_1}h^2$, 
although this bound on WIMP overproduction occurs at over an order of 
magnitude higher values: in Fig. \ref{fig:Oh2}, $f_a\alt 10^{14}$ GeV
occurs from just overproduction of the WIMP component of dark matter.
\begin{figure}[tbp]
\begin{center}
\includegraphics[height=0.35\textheight]{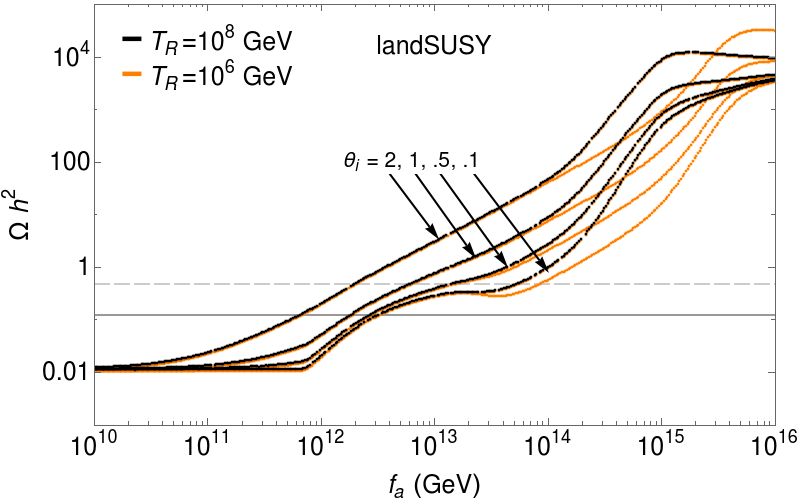}
\caption{
Relic density of total axion plus higgsino-like WIMP DM versus $f_a$.
Results here are for $\theta_i=0.1,\ 0.5,\ 1$ and $2$ and for
$\theta_i=1$ but with $T_R=10^6$ and $10^8$ GeV.
\label{fig:Oh2_2}}
\end{center}
\end{figure}

In Fig. \ref{fig:Oh2_2}, we show the total mixed WIMP plus axion dark matter 
abundance but this time assuming $T_R=10^6$ GeV and $10^8$ GeV with 
$\theta_i=0.1,\ 0.5,\ 1$ and 2.
For $T_R\agt 10^9$ GeV, thermal production and 
late decay of gravitinos can lead to conflict with bounds from late-decaying 
neutral particles in the early universe: in this case, the gravitino 
problem\cite{Kawasaki:2008qe,Jedamzik:2006xz}. 
We see that for different $\theta_i$ values the upper limit on
$f_a$ can move around by typically an order of magnitude: 
nonetheless, an upper bound on $f_a$ from overproduction of dark matter 
should obtain which is still much less than the the string/GUT 
scale. We also show variation in the $a-\tz_1$ dark matter relic density
versus varying $T_R$. For the DFSZ axion model, the axino and saxion production
rates in the early universe hardly depend on $T_R$\cite{Bae:2011jb} 
(unlike the case of the KSVZ axion model\cite{Brandenburg:2004du}). 
Some variation in relic density is seen for 
$f_a\agt 10^{14}$ GeV where gravitino production, 
which does depend on $T_R$\cite{Pradler:2006qh},
becomes important and augments the non-thermal WIMP abundance.

\subsection{PQ scale from the landscape}
\label{ssec:PQland}

In this Subsection, we investigate whether landscape considerations can
determine the magnitude of the PQ scale $f_a$.
We assume an $n=1$ statistical draw towards large soft terms $-A_f$ 
which in turn leads to large PQ scales along the lines of 
Fig. \ref{fig:fa} where the PQ scale is related to the breakdown of 
supersymmetry. 
For our landscape benchmark point landSUSY, the magnitude of $f_a$ is 
determined by the quartic soft term $A_f$. 
However, since $-A_f$ is not connected with EWSB, 
then it need not be susceptible to the same bounds on MSSM soft terms
that emerge from requiring an appropriate breakdown of electroweak 
symmetry with independent contributions to $m_{weak}$ not more than a factor of
a few from its value $m_{weak}\simeq 100$ GeV. 
Instead, the PQ scale $f_a$ is intimately related to the production of 
both axion dark matter and (natural) higgsino-like WIMP dark matter.

Since we are working within a multiverse scenario wherein each pocket universe
may have different laws of physics, and the multiverse is an expression of
the universe emerging from a spacetime continuum characterized by 
eternal inflation, then of course inflationary cosmology is an essential
component of our overall scheme. In inflationary cosmology, the universe
has an early exponential expansion phase which drives the universe to 
flatness, which requires an overall energy density teetering on the boundary
between an open or a closed universe. Such a universe is characterized by the
overall energy density lying at its critical closure density:
\bea
\rho &=&\rho_c=3H_0^2/8\pi G_N\ \ {\rm or}\ \ \Omega\equiv \rho/\rho_c =1 \\
{\rm with}\ \ \ \Omega &\equiv & \Omega_B+\Omega_{rad}+\Omega_{DM}+\Omega_\Lambda+\Omega_{curv}
\eea
and where $\Omega_{curv}=0$ for an inflationary universe which gives rise to 
a flat geometry.
For our pocket universe, the measured value of the Hubble constant 
is $H_0=100 h$ km/s/Mpc with $h=0.678\pm 0.009$ but for other pocket
universes then $H_0$ will be different depending on the 
various constituencies. We will adopt as usual $\rho_B/\rho_\gamma$ equal 
to the value of our universe since we are assuming a ``friendly'' 
fertile patch of the multiverse where the SM remains as the low energy 
effective theory.\footnote{
Anthropic arguments usually depend on a so-called 
``friendly'' landscape
wherein one focuses on most parameters asuming their SM values so as
to retain predictivity\cite{ArkaniHamed:2005yv}. Sometimes these
are called {\it fertile patches} of the landscape of vacua since they
should lead to the standard cosmological and particle physics models
aside from just the few mass scales which may scan in the multiverse.}
Thus, in the fertile patch of multiverse assumed here, only
$\Lambda$, $m_{soft}$ and $\theta_i$ are assumed to scan.
The scanning of the soft term $A_f$ sets the value of $f_a$ 
for otherwise fixed values of scalar masses as expected for our 
landSUSY benchmark point: {\it i.e.} we assume a common value of all 
scalar masses $m_0(1,2)=m_X=m_Y\equiv m_{3/2}$. We allow smaller
values of $m_0(3)$ as occurs in the mini-landscape picture wherein
third generation fields lie on the bulk of the compactified orbifold
whilst first/second generation fields lie near orbifold fixed 
points\cite{Nilles:2014owa}.

\begin{figure}[tbp]
\begin{center}
\includegraphics[height=0.23\textheight]{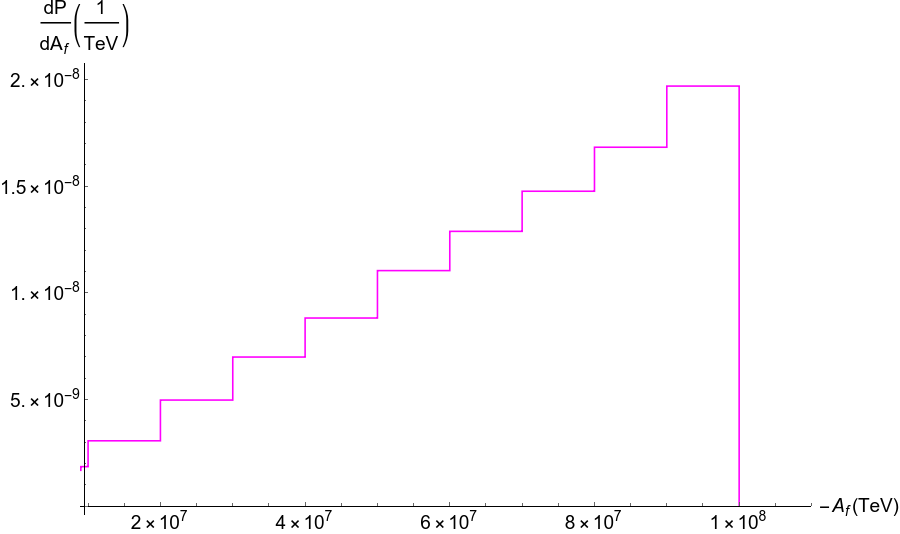}
\includegraphics[height=0.23\textheight]{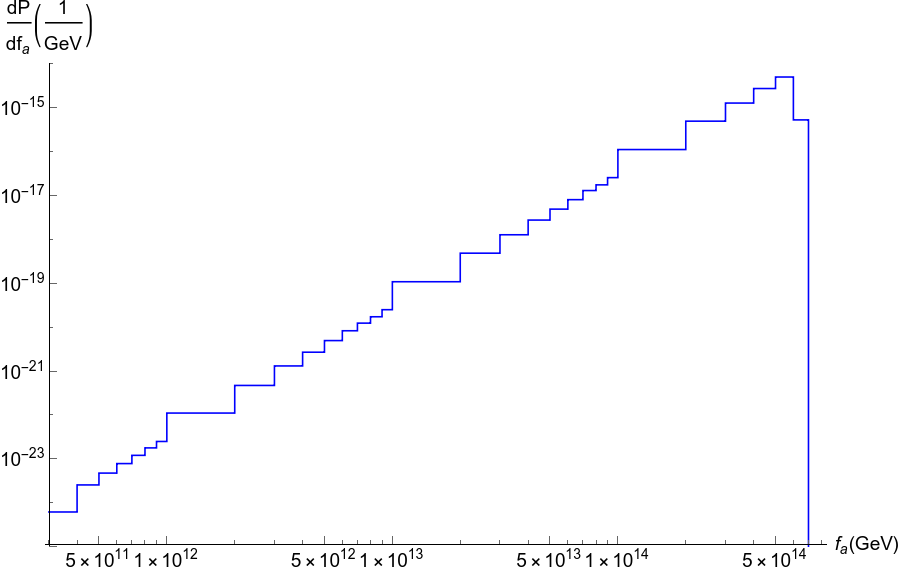}
\caption{In {\it a}), we show the assumed distribution of soft SUSY 
breaking term $-A_f$ from an $n=1$ statistical pull from the landscape.
In {\it b}), we show the corresponding probability distribution in $f_a$.
\label{fig:Af_fa}}
\end{center}
\end{figure}

The generated probability distribution for $-A_f$ is shown in 
Fig.~\ref{fig:Af_fa}{\it a}), which is seen to rise linearly as expected.
For a given value of $A_f$, then the value of $f_a$ is determined by
the minimization conditions arising from Eq. \ref{eq:VhyCCK}. 
In frame {\it b}), we show the derived distribution $dP/df_a$. 
Here, the probability distribution is seen to favor the highest
values of $f_a$ possible, which would be generated from very large
values of $-A_f$. 

At this point, our prior distribution for $f_a$ is set, but we will also need
some selection criterion to avoid $f_a$ exploding up to huge values, 
leading to perhaps a gross overproduction of dark matter. 
Thus, the question now is: how much dark matter is too much dark matter
for our fertile patch of pocket universes within the greater multiverse?
Some of the papers of Subsec.~\ref{ssec:review} have entertained values of 
$\rho_{DM}/\rho_B$ as high as 25-100. 

For illustrative purposes, we will consider the effect of 
limiting pocket universes to a modest bound of four times greater 
dark matter density than in our universe: suppose $\Omega_{\rm DM}h^2\alt 0.48$. 
Such a bound would saturate the case
where we maintain our measured value of $\rho_c$ but allow the dark matter
abundance to nearly saturate $\rho_c$ at the expense of a dark energy component.
Such models were commonly contemplated before the discovery of 
a non-zero dark energy component. 

\begin{figure}
\centering
\begin{subfigure}[t]{0.46\textwidth}
  \centering
  \includegraphics[width=1.1\linewidth]{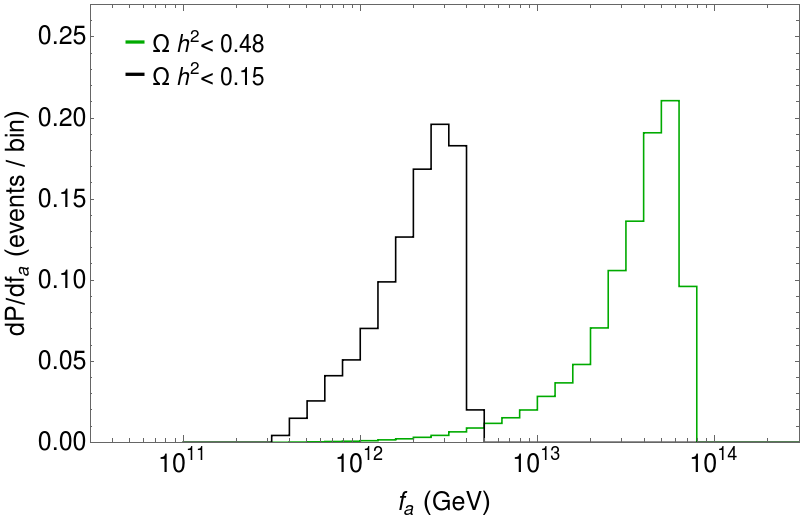}
  \caption{}
  \label{fig:dPdfa}
\end{subfigure}%
\quad \quad
\begin{subfigure}[t]{0.44\textwidth}
  \centering
  \includegraphics[width=1.1\linewidth]{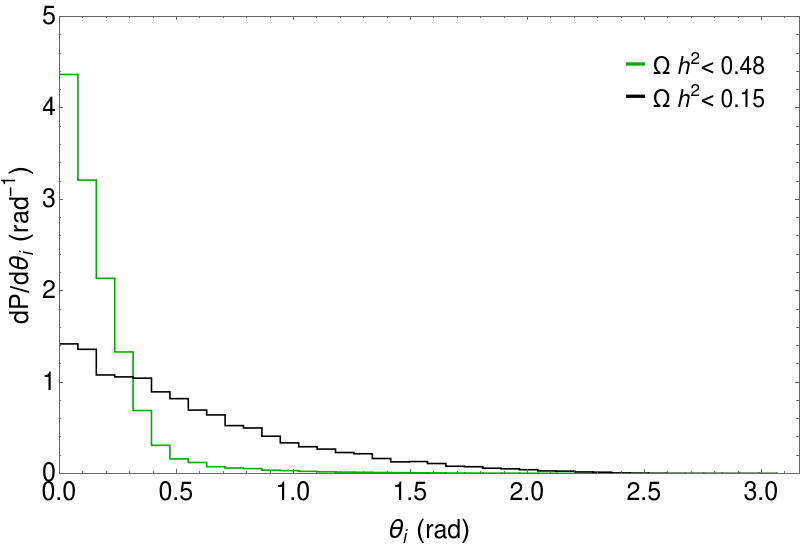}
  \caption{}
  \label{fig:dPdti}
\end{subfigure}
\caption{Probability distribution in {\it a}) $f_a$ and {\it b}) $\theta_i$
assuming an $n=1$ statistical pull on the 
soft SUSY breaking term $-A_f$ from the landscape and requiring
no more than a factor four more DM (green) or else $\Omega_{\rm DM}h^2\le 0.15$
(black).}
\label{fig:d2}
\end{figure}

In Fig.~\ref{fig:dPdfa}), we show the resulting probability distribution
$dP/df_a$ which results from an $n=1$ draw on $-A_f$ coupled to an
anthropic/cosmological selection bound $\Omega_{a\tz_1}h^2<0.48$ 
(green curve).
Even with our proposed modest selection bound, we see that the value of
$f_a$ is driven to its nearly maximal value such as to avoid overproduction 
of dark matter. From the plot, we would expect that a value of 
$f_a\sim 10^{14}$ GeV or only somewhat lower, with a rather sharp cutoff
$f_a\alt 8\times 10^{13}$ GeV. For comparison, we also show the black
histogram where we instead require that the upper bound on dark matter 
abundance is only slightly beyond our measured value: $\Omega_{\rm DM}<0.15$.
This case would prefer $f_a\sim 5\times 10^{12}$ GeV.

\begin{figure}[tbp]
\begin{center}
\includegraphics[height=0.43\textheight]{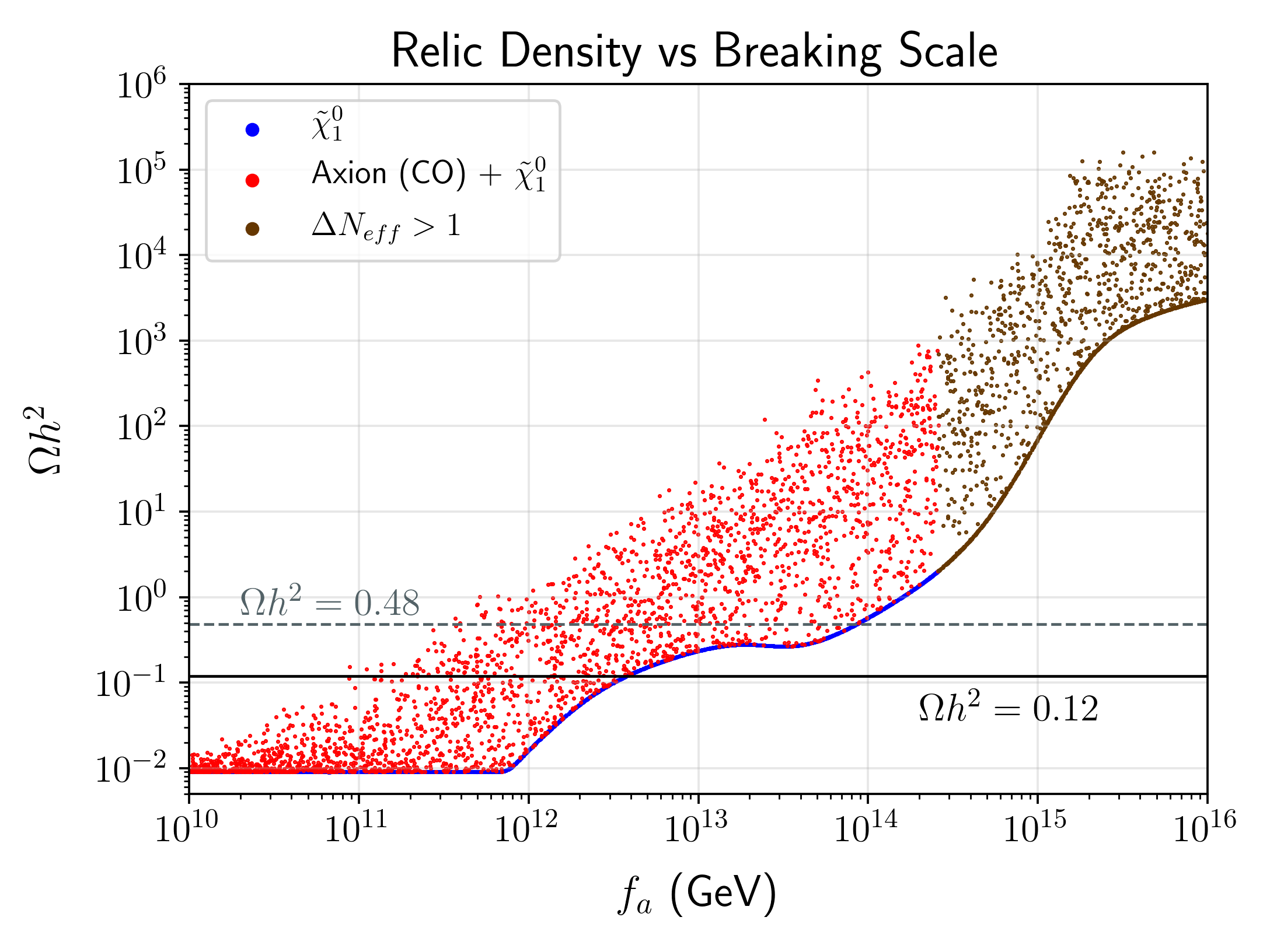}
\caption{
Range of relic density values for axion and higgsino-like WIMP 
dark matter versus $f_a$ from uniform scan over $\theta_i$ with 
$m_{\ta}=m_s=16$ TeV in the SUSY DFSZ axion model. 
(The blue points lie along the lower boundary of plotted points.)
\label{fig:dfsz}}
\end{center}
\end{figure}

Let us compare the results of Fig. \ref{fig:dPdfa}) with those of
Fig. \ref{fig:dfsz} which shows the allowed mixed axion-WIMP 
dark matter abundance
for our landSUSY benchmark point in the generic SUSY DFSZ axion model 
while scanning uniformly over $\theta_i$ and uniformly over $\log (f_a)$.
From Fig. \ref{fig:dfsz}, we see that for $f_a\sim 10^{13}-10^{14}$ GeV,
we are already overproducing dark matter compared to our universe 
with $\Omega_{\rm DM}h^2=0.12$. There is only a miniscule probability to
obtain from Fig. \ref{fig:dPdfa}) $f_a$ values low enough to match the 
measured value, which occurs for $f_a\sim 10^{11}$ -- $\sim 4\times 10^{12}$
GeV. In the references from Subsec.~\ref{ssec:review}, large values of
$f_a\sim 10^{14}-10^{16}$ GeV could be compensated for by selecting on 
small values of $\theta_i$. 
For our case of natural mixed axion-WIMP dark matter, this compensation is not
permitted because large $f_a$ also leads to large (non-thermal) 
overproduction of WIMP dark matter via delayed axino and saxion decays 
in the early universe. From Fig. \ref{fig:dPdfa}, we would expect
that if the landscape is involved in determining the PQ scale $f_a$, 
then its value should be very near the maximally allowed abundance of DM 
in pocket universes such as to allow observers to exist. 
But it is hard to believe that our pocket universe's value of dark matter 
abundance is nearly anthropically maximal (as depicted by the black curve of
Fig. \ref{fig:dPdfa}). 
\begin{figure}[tbp]
\begin{center}
\includegraphics[height=0.36\textheight]{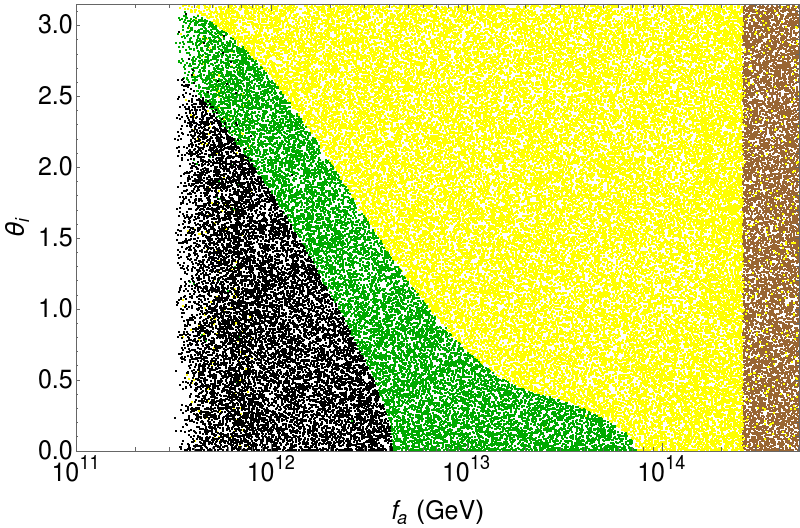}
\caption{Allowed and disallowed (yellow) points in the $f_a$ vs. 
$\theta_i$ plane assuming a modest selection bound of $\Omega_{\rm DM}h^2<0.48$
(green) and $\Omega_{\rm DM}h^2<0.15$ (black).
\label{fig:fa_ti}}
\end{center}
\end{figure}

In Fig.~\ref{fig:dPdti}), we show the corresponding distribution 
$dP/d\theta_i$ from the $n=1$ pull on soft terms coupled with
our modest  anthropic veto that $\Omega_{\rm DM}h^2<0.48$. The plot shows
a probability that $\theta_i$ is peaked around its smallest allowed values.
This is easy to understand in that while the landscape prior 
strongly favors large values of $f_a$, from Eq. \ref{eq:axionCO} we see that 
overproduction of axions can be avoided by selecting only those vacua 
with correspondingly tiny values of $\theta_i$. 
This effect is easily understood from Fig. \ref{fig:fa_ti} where we show
regions of the $\theta_i$ vs. $f_a$ plane for our landSUSY benchmark point
which lead to $\Omega_{\rm DM}h^2<0.48$ (green points) or $\Omega_{\rm DM}h^2>0.48$
(yellow points). The brown points denote where also $\Delta N_{eff}>1$.
From the figure, we see that for large $f_a\sim 8 \times 10^{13}$ GeV, only a small
range of $\theta_i$ allows for non-overproduction of dark matter.
And once $f_a\agt 8 \times 10^{13}$ GeV, then no value of $\theta_i$ is possible which
allows one to avoid DM overproduction. 

%
%

\section{Prediction of PQ scale from generic SUSY DFSZ 
axion model with uniform scan on $\theta_i$}
\label{sec:dfsz}

In Sec. \ref{sec:Z24R}, we adopted a particular gravity-safe SUSY axion model
based on a $\mathbb{Z}_{24}^R$ discrete $R$-symmetry and the hyCCK 
superpotential Eq. \ref{eq:WhyCCK} to show that a statistical draw 
towards large soft terms also yields a draw to large PQ breaking scale $f_a$. 
The value of $f_a$ gains an upper bound by requiring no overproduction of 
dark matter. 
For the modest assumption of less than a factor four times the measured
abundance of dark matter, then we found $f_a\sim 10^{14}$ GeV 
which is well below the values expected from pre-landscape string theory
but which typically leads to much more dark matter production than we 
observe in our universe.

In this Section, we try to be more general by eschewing a particular
SUSY axion model and instead assume a generic SUSY DFSZ axion 
model\cite{dfsz2} where $f_a$ is an input instead of an output parameter. 
In this case, we will adopt  a uniform distribution in $\theta_i$ 
in accord with expectations from the landscape, but then require that the
dark matter abundance lie at its measured value: $\Omega_{a\tz_1}h^2=0.12$.
From this, we can then determine the necessary value of $f_a$ such that,
for scanned values of $m_{\ta}$, $m_s$ and $m_{3/2}$, the measured
abundance of mixed axion-neutralino dark matter is obtained.
We will scan uniformly over each of $m_{\ta},\ m_s$ and $m_{3/2}:1-50$ TeV. 
\begin{figure}[tbp]
\begin{center}
\includegraphics[height=0.3\textheight]{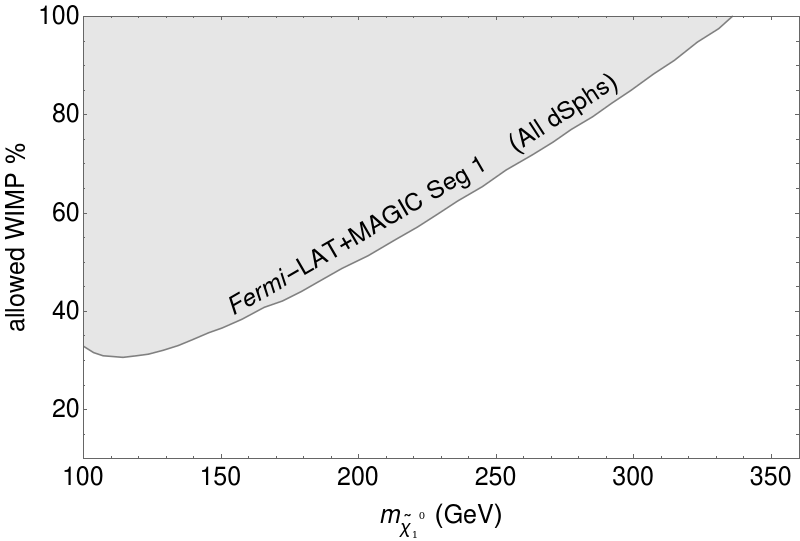}
\caption{Percent of neutralino dark matter contributing to total dark matter
vs. $m_{\tz_1}\simeq \mu$ compared to recent limits from Fermi-LAT+MAGIC bounds
on gamma rays from dwarf spheroidal galaxies.
\label{fig:percent}}
\end{center}
\end{figure}

For a SUSY benchmark point within a two-component dark matter framework, 
direct and indirect WIMP dark matter searches can put a stringent 
upper limits upon the neutralino density which are more severe 
than the measured value, $\Omega_{\rm DM}h^2=0.12$. 
In many cases, indirect DM detection (IDD) offers the most contraining limits 
on the non-thermal, decay-produced neutralinos for models with thermally 
underproduced higgsino-like neutralinos. 
In natural SUSY models from the $n=1$ landscape, thermally produced neutralinos 
typically make up 5-20\% of the total CDM density which renders them safe from 
Fermi-LAT+MAGIC\cite{fermi} limits on overproduction of gamma rays in 
dwarf spheroidal galaxies\cite{axion}. 
In Fig. \ref{fig:percent}, we show the allowed percentage of WIMP dark matter 
compared to $m_{\tz_1}$ along with the Fermi-LAT+MAGIC IDD limit for our landSUSY benchmark
point. If we increase $\mu\sim 340$ GeV, then $m_{\tz_1}\sim 340$ GeV and all generated points would be Fermi-LAT+MAGIC allowed. The gray-shaded region shows the excluded WIMP composition for all landSUSY points within a good approximation.

\begin{figure}[tbp]
\begin{center}
\includegraphics[height=0.24\textheight]{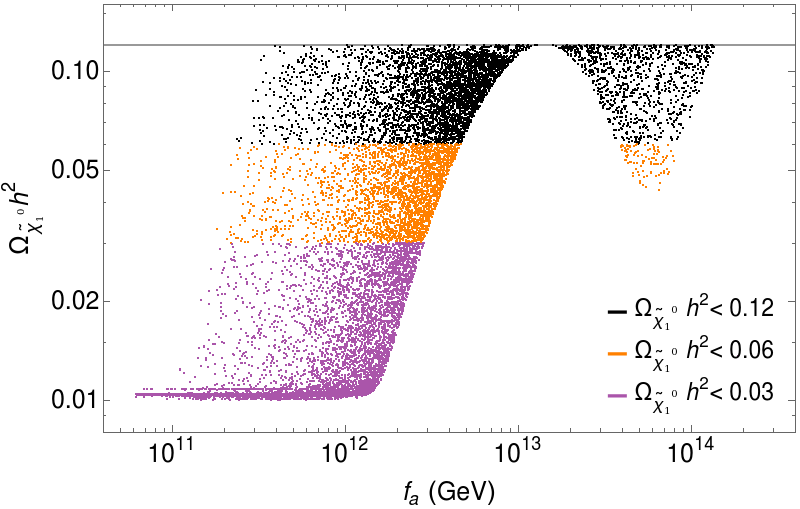}
\includegraphics[height=0.24\textheight]{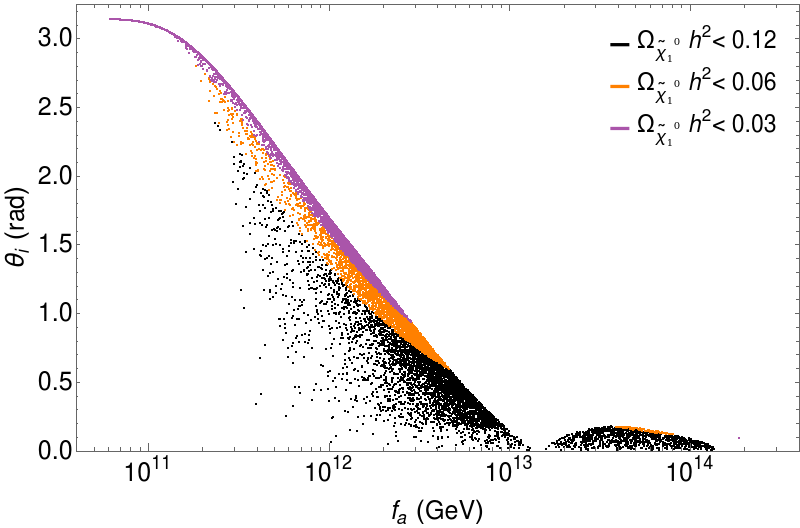}\\
\includegraphics[height=0.24\textheight]{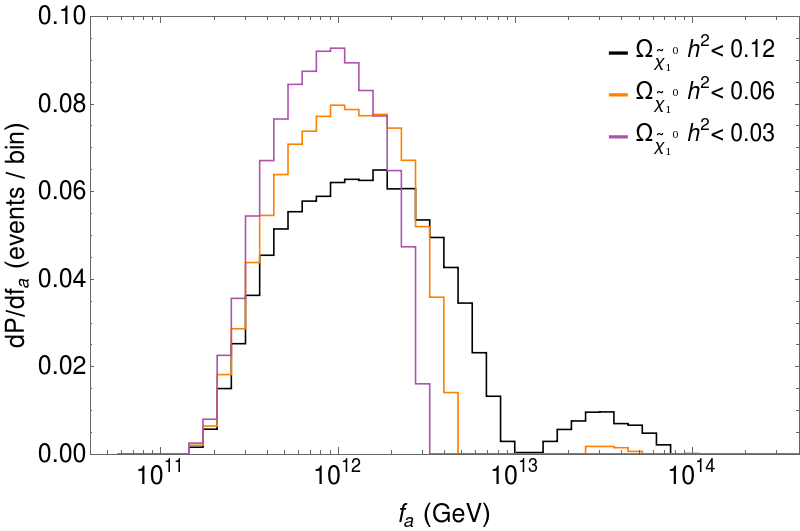}
\includegraphics[height=0.24\textheight]{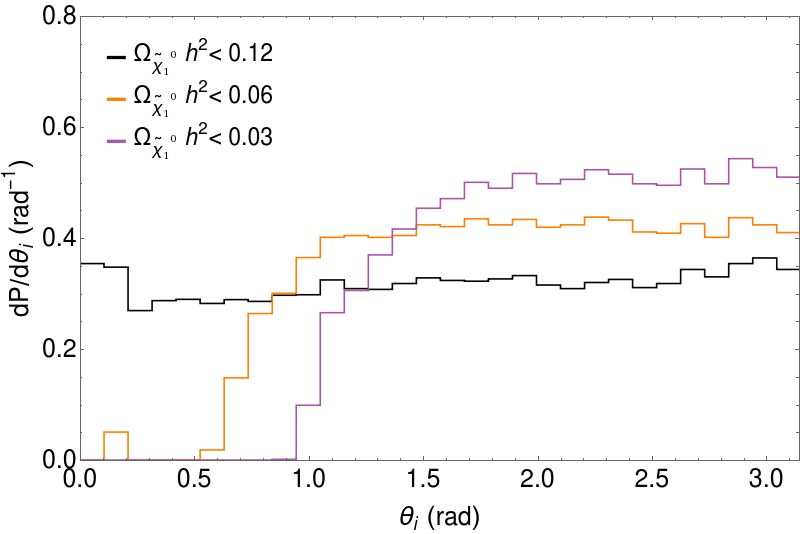}
\caption{In {\it a}), we plot the value of $\Omega_{\tz_1}h^2$ versus
$f_a$ from a uniform scan over $\theta_i: 0\to 3.14$ 
(and $m_{\ta}$, $m_s$ and $m_{3/2}$). In {\it b}), we show the corresponding
correlation of $\theta_i$ vs. $f_a$. 
In {\it c}), we show the ensuing probability distribution for 
$f_a$.
In {\it d}), we show the probability distribution in $\theta_i$ after
selection effects. In all the frames, we require the total abundance of DM to equal its measured value: $\Omega_{a\tz_1}h^2=0.12$.
\label{fig:dfszscan}}
\end{center}
\end{figure}

In Fig. \ref{fig:dfszscan}, we work within the SUSY DFSZ axion model 
using again our landSUSY benchmark point but with input parameters
$m_s,\ \theta_s,\ \theta_i,\ f_a,\ T_R$ and $m_{3/2}$. 
Here, we fix $T_R=10^7$ GeV and $\theta_s=1$ but allow
$m_s$, $m_{\ta}$ and $m_{3/2}$ to scan over the range given above 
with a uniform scan on $\theta_i$ and a log prior scan on $f_a$. We only accept solutions with $\Omega_{a\tz_1}h^2=0.12$. We show the parameter space with augmented neutralino densities $\Omega_{\tz_1}h^2 < 0.12, 0.06$ and 0.03 with black, orange and purple colors respectively. We impose an upper limit on $\theta_i$ ($\theta_i < 3.14$) so that the highly fine-tuned region $\theta_i \simeq \pi$ is not present in our analysis.

In Fig. \ref{fig:dfszscan} frame {\it a}), the resulting abundance of neutralino dark matter is shown
while the remainder of DM is made of DFSZ axions. The horizontal line
around $f_a \lesssim 10^{11}$ GeV is just the expected thermal abundance of
200 GeV higgsino-like WIMP dark matter. For higher $f_a > 10^{11}$
GeV, then non-thermal LSP production begins to occur where axinos
can be produced in the early universe and decay to LSPs after 
neutralino freeze-out. There is a gap around $f_a\sim 10^{13}$ GeV
where axino decays are still contributing to the neutralino density.
For $f_a\sim 10^{14}$ GeV, points again become allowed due to 
diminished thermal production of axinos in the early universe.
The WIMP abundance increases for $f_a\agt 10^{14}$ GeV  due to increasing CO-production
of saxions which then decay (in part) to WIMPs\cite{dfsz2}.
An upper limit of $f_a\alt 2\times 10^{14}$ GeV ensues 
in this case since for large $f_a$, the DM is always overproduced. 
There is no conflict here with Fig. \ref{fig:dfsz} since in this case
with random values of $m_s$ and $m_{\ta}$, then 
relative axino and saxion production and decay rates can vary
which leads to allowed points for $f_a\sim 10^{14}$ GeV.

The corresponding correlation of $\theta_i$ with the required value of
$f_a$ to make $\Omega_{a\tz_1}=0.12$ is shown in frame {\it b}). 
Here, large values of $\theta_i$ are correlated with low values of $f_a$ 
to boost the axion production to gain accord with the measured relic abundance.
For very large $f_a$, then consequently small values of $\theta_i$
are required to allow for $\Omega_{a\tz_1}h^2=0.12$.

In frame {\it c}), we show the resulting probability distribution
$dP/df_a$ versus $f_a$. Including all points with the measured abundance, 
then one obtains the black histogram
which peaks around $f_a\sim 2 \times 10^{12}$ GeV but with a tail extending to over
$10^{14}$ GeV. For the cases in which neutralino makes less than half of the measured DM density, the peak shifts to lower values of $f_a$ (orange and purple histograms) with a small probability at high $f_a$.

In frame {\it d}), we show the probability distribution $dP/d\theta_i$
for the three cases considered. Here, the black histogram is almost uniform across its
range whilst the orange histogram displays a gap at small $\theta_i$ 
where no allowed solutions occur. 
The high $f_a$ region does not show up for $\Omega_{\tz_1}h^2 \lesssim 0.04$ and 
$\theta_i$ can only take values greater than $\sim$1 when the neutralino 
makes less than 25\% of the total DM abundance (purple).

\section{Conclusions}
\label{sec:conclude}

In this paper we have sought to answer the question: is the magnitude of the
PQ scale $f_a$ set by the landscape, or by something else?
To address this question, we have adopted the scenario 
advocated by Douglas wherein the soft terms are statistically favored
by a prior distribution $m_{soft}^{2n_F+n_D-1}$ and where we take the 
value $n=2n_F+n_D-1=1$ ({\it i.e.} a linear distribution favoring large soft
SUSY breaking terms).
Along with this prior distribution, we invoke a selection criteria that 
vetos models with inappropriate EW breaking (CCB minima or no EWSB) 
and vetos models with contributions to the weak scale $\agt 4$ 
(corresponding to $\Delta_{\rm EW}>30$) in accord with nuclear physics constraints
derived by Agrawal {\it et al.} on anthropically allowed values 
for the weak scale. Such an approach receives support in that previously 
it has been shown that $n=1$ (or 2)
has a most probable Higgs mass of $m_h\simeq 125$ GeV whilst lifting 
sparticle masses beyond the reach of Run 2 of the LHC. 

We implement this approach within a highly motivated SUSY axion model labeled
as hyCCK. We feel this is an improvement upon previous work in that the model
1. stabilizes the weak scale via SUSY, 2. solves the strong CP problem
via a gravity-safe $\mathbb{Z}_{24}^R$ discrete symmetry which gives rise
to an accidental, approximate global $U(1)_{PQ}$, 3. solves the SUSY 
$\mu$ problem\cite{mu} via the presence of a 
Kim-Nilles superpotential\cite{KN} and 4.
explains $R$-parity conservation and proton stability as consequences of the
more fundamental $\mathbb{Z}_{24}^R$ which could arise from compactification
of extra space dimensions in string theory.
By choosing a MSSM benchmark point in accord with $n=1$ landscape predictions
and allowing for the PQ soft term $-A_f$ to scan linearly and to 
set the magnitude of the PQ scale $f_a$, then we find that as large as possible
values of $f_a$ are statistically prefered. 
In this approach, typically both WIMP and axion dark matter are overproduced.
Axions are overproduced via vacuum misalignment at $f_a\gg 10^{12}$ GeV 
unless the large value of $f_a$ is compensated for by a small value of
$\theta_i$. However, since WIMPs are also overproduced at large $f_a$, due
to axino and saxion production coupled with delayed decays to SUSY LSPs after
neutralino freeze-out ({\it i.e.} non-thermal WIMP production), then even
with small $\theta_i$ one cannot avoid overproduction of mixed axion-WIMP
dark matter. From this approach, using even a modest DM overproduction bound
of a factor four, then there is only a tiny probability to gain the measured
value of dark matter density in our own pocket universe. Thus, the answer to the question posed in the title is: No, in our well-motivated 
landscape SUSY model based upon gravity-safe, electroweak natural hyCCK 
SUSY axion model, the magnitude of the PQ scale is highly unlikely to be set 
by the landscape.

Instead, an alternative but perhaps underappreciated mechanism is available
to set the magnitude of the PQ scale. 
This is that in generic supergravity models with hidden sector 
SUGRA breaking via the superHiggs mechanism, then soft terms arise from
SUGRA breaking with magnitudes of order the gravitino mass $m_{3/2}$.
For a well-specified hidden sector, then the soft terms are all calculable 
and correlated.
For our landscape SUSY model with $m_0(1,2)\sim m_{3/2}$, we would also
expect $-A_f\sim m_{3/2}\sim 10-100$ TeV. This places us from Fig. \ref{fig:fa}
into the zone where $f_a\sim 10^{11}-10^{12}$ GeV which is the sweet spot
for generating a thermal underabundance of higgsino-like WIMP dark matter
but with mainly SUSY DFSZ axion dark matter.  

In Sec.~\ref{sec:dfsz}, we implemented instead a uniform scan over 
$\theta_i$ and a scan over independent $m_{\ta}$, $m_s$ and $m_{3/2}$ values 
with the corresponding $f_a$ value such that 
$\Omega_{a\tz_1}h^2=0.12$. 
In this case, we are pushed to the cosmological sweet spot where
$f_a\simeq 10^{11}-10^{13}$ GeV.

From our overall approach, we understand why WIMPs haven't yet been detected: 
it is because they make up only a small portion of the total dark matter. 
In addition, even though the non-thermal decay-production of WIMPs 
is allowed for, more than 40\% of the points in 
Fig.~\ref{fig:dfszscan}{\it a}) (Sec.~\ref{sec:dfsz}) 
still have the same neutralino abundance as the thermally-produced value.
Nonetheless, WIMP discovery should be possible at multi-ton noble liquid 
direct WIMP detection experiments\cite{wimp}. 
Axions-- while likely to make up the bulk of dark matter-- 
are much more difficult to detect
since the presence of higgsinos in their $g_{a\gamma\gamma}$ coupling
diagram suppresses their detection rate compared to KSVZ or 
non-SUSY DFSZ axion models\cite{axion}. Meanwhile, sparticle detection
may be possible at HL-LHC via the soft OS dilepton+jet+MET channel 
which arises from direct higgsino pair production\cite{SDLJMET}. Detection of 
gluinos, top squarks or winos might be possible at HL-LHC if we are lucky, but 
otherwise may have to await construction of higher energy $pp$ 
colliders\cite{jamie}.

{\it Acknowledgements:} 
This work was supported in part by the US Department of Energy, Office
of High Energy Physics. The computing for this project was performed at the OU Supercomputing Center
for Education \& Research (OSCER) at the University of Oklahoma (OU).


%
\end{document}